\documentclass[12pt]{article}
\input{tcilatex}
\input tcilatex

\begin{document}

\title{Looking beyond the Thermal Horizon: Hidden Symmetries in Chiral Models}
\author{B. Schroer and H.-W. Wiesbrock \\
%EndAName
FB Physik der FU Berlin\\
Arnimallee 14, 14193 Berlin, Germany\\
schroer@physik.fu-berlin.de}
\date{November, 1998}
\maketitle

\begin{abstract}
In thermal states of chiral theories, as recently investigated by H.-J.
Borchers and J. Yngvason, there exists a rich group of hidden symmetries.
Here we show that this leads to a radical converse of of the Hawking-Unruh
observation in the following sense. The algebraic commutant of the algebra
associated with a (heat bath) thermal chiral system can be used to reprocess
the thermal system into a ground state system on a larger algebra with a
larger localization space-time. This happens in such a way that the original
system appears as a kind of generalized Unruh restriction of the ground
state sytem and the thermal commutant as being transmutated into newly
created ``virgin space-time region'' behind a horizon. The related concepts
of a ``chiral conformal core'' and the possibility of a ``blow-up'' of the
latter suggest interesting ideas on localization of degrees of freedom with
possible repercussion on how to define quantum entropy of localized matter
content in Local Quantum Physics.
\end{abstract}

\section{Introduction}

In a previous paper we defined and illustrated the concept of a ``hidden
symmetry'' \cite{S-W1}. We remind the reader that we use the terminology
``hidden'' for such symmetries of local quantum physics (LQP) which cannot
be seen and understood by (Lagrangian-, functional integral-) quantization.
They simply do not exist on a classical level nor in a euclidean functional
integral representation, and hence cannot be transposed into the
noncommutative realm by e.g. adapting ( via quantization\footnote{%
If a quantization approach could lead to nonperturbative mathematically
controllable real time local quantum physics, then these symmetries would
show up in modular properties of the local algebras generated by the fields.
But as everybody should know, we are (despite intense efforts for more than
40 years) still far away from such a goal, and it appears presently much
more plausible that this will be achieved by modular methods \cite{S-W2}
instead of quantization.}) Noether's theorem. One rather needs the power of
the modular theory, which is totally characteristic of the noncommutative
aspects of LQP (\textbf{L}ocal \textbf{Q}uantum \textbf{P}hysics), and which
in addition to the classical space-time diffeomorphisms (Poincar\'{e}- or
Conformal- automorphisms) of Minkowski space-time also reveals a host of
non-Noetherian ''Hidden Symmetries'' of modular origin. The latter escape
the quantization approach to LQP (see the previous footnote), whereas the
modular approach to LQP shows both kinds of symmetries; in fact the hidden
symmetries may be considered in some sense as being the generalization of
the diffeomorphisms of chiral conformal theory to higher dimensions \cite
{S-W1}.

There are two mechanisms which reveal such hidden structures: the modular
theories of (multiply) localized double cone algebras together with suitably
chosen states, and the symmetries constructed from modular inclusions and
modular intersections. The first mechanism might lead to a ''fuzzy'' action%
{\Large \ }of the modular group for which the double cone situation for the
free massive theory is the simplest illustration. It is believed that such
modular actions become asymptotically pointlike near the horizon (boundary)
of the localization region \cite{S-W1}. We will use the word hidden in this
paper exclusively for the second inclusion/intersection mechanism which
leads to (partially) geometric transformations of localization regions (i.e.
of the net) which cannot be implemented as point transformations. For
details see below and \cite{S-W1}. 

Modular inclusions/intersections of von Neumann algebras are significantly
different from the well-studied V. Jones inclusions (subfactors). Whereas by
definition the latter have conditional expectations onto the smaller algebra
(``shallow'' inclusions), the former, as a result of their modular
structure, cannot have conditional expectations (``deep'' inclusions). In
passing we mention that there is a third type of deep inclusion which also
has no conditional expectation and is not modular;{\large \ }it is the
so-called split inclusion which is related to a strong form of statistical
independence in LQP \cite{Su}.

It is well-known that the V. Jones inclusions lead to vast generalizations
of compact group symmetry. On the other hand modular
inclusions/intersections have only been around for some years and very
little is known about their consequences except that noncompact groups (as
the two parametric translation/dilation group as well as SL(2,\textbf{R}))
emerge already from the simplest examples.{\Large \ }Although modular (hsm)
standard inclusions are known to be in one-to-one correspondence to strongly
additive chiral quantum fields (and thereby there are good reasons for
expecting to be able to classify them), no results and even no efforts in
this direction is known today.{\Large \ } These groups act in a natural way
on the two original algebras and their intersection and generate a
space-time indexed net. The considerations in this paper lead to an increase
in knowledge about modular inclusions/intersections.

In our previous work we commented among other things on a very revealing
chiral illustration of such a situation which was found by Borchers and
Yngvason \cite{Bo/Yng}. In this note we use their example as a starting
point for a furthergoing analysis of hidden symmetries in low-dimensional
theories.

As was shown by those authors, it might happen that the modular group acts
only geometrically on a certain subregion. We show that in their case this
group acts geometrically in a much larger region! Moreover there is a much
richer ``hidden symmetry'' behind. In fact it turns out that in addition to
the hidden symmetry generated by the two-parametric translation/dilation
group \cite{Bo/Yng} (with the dilation part being hidden), there is a hidden
SL(2,\textbf{R}) M\"{o}bius symmetry with a differently localized net which
is partially local relative to the old thermal net. In fact the thermal
``shadow world'' associated with a heat bath state becomes converted into a
piece of real space-time which is to be localized behind the horizon! This
is an inversion from the Hawking-Unruh situation in a two-fold sense: not
only is the thermalization through localizing (Unruh) horizons undone and
the thermal region returned as a piece of Minkowski space to the vacuum
world (in analogy to the Unruh effect \cite{S/S-V}), but even the ``shadow
world'' (the apparently non-geometric thermal commutant) of a standard 
\textit{heat-bath thermal situation} becomes converted into ``virgin \textit{%
space-time}'' \textit{behind the horizon}, thus giving additional weight to
some speculative remarks of Jaekel \cite{Ja}. Although we can presently
illustrate this idea only in the chiral light ray setting, we consider the
potential possibility of a space time interpretation of heat bath
temperature, and a still unknown but expected ensuing conversion of a (still
missing) notion of quantum entropy as sufficiently intriguing in order to
justify publication.

Using our recent ideas on the modular origin of the chiral diffeomorphism
group which we illustrated with some computations on the Weyl-algebra \cite
{S-W1}, it could be interesting to extend also these ideas about a modular
origin of chiral diffeomorphism groups to the thermal setting. Leaving this
for future investigations, we make some remarks on the existence of hidden
SL(2,\textbf{R})\ symmetries in d=1+1 massive theories and we conclude with
some comments on higher dimensional theories. In all cases the modular
method is used to extend the hidden action of translation/dilation subgroup 
\cite{S-W1} to a hidden SL(2,\textbf{R}) action.

\section{Beyond the Borchers-Yngvason Results}

Consider a chiral quantum field theory given by a local net of observables.
The Bisognano-Wichmann property holds for the net, as it was in general
shown by \cite{B/G/L, Fr/Ga}. For simplicity we also assume strong
additivity or equivalently, Haag duality on the line, see \cite{G/L/W}. This
can always be achieved by passing to the dual net. With Borchers and
Yngvason we take a thermal state on the net w.r.t. the lightlike
translations. Let $\mathcal{N}$ denote the observable algebra to the
half-line $[0,\infty [,$ then Borchers and Yngvason showed that the modular
group of $\mathcal{N}$ w.r.t. the thermal state transforms local algebras on
the half-line geometrically (see below). The global thermal algebra is known
to be of hyperfinite von Neumann type III$_{1}$ and hence inequivalent to a
ground state representation. This is the prerequisite for an Unruh-like
interpretation in terms of a ground state representation thermalized by
local restriction.

We will show that this group also acts geometrically on the whole line and
that even more, there is a ''hidden'' conformal group SU(2,\textbf{R})/Z$%
_{2} $ acting geometrically on intervals, but not pointwise in the sense of
spatial automorphisms..

For describing our observation, let us first recall the setting of Borchers
and Yngvason. Denote by $\mathcal{A}$ the quasilocal algebra of the chiral
model, by $\Omega _{+}$ the GNS-vector to the thermal state w.r.t. the
unitary group $U(a)$ of lightlike translations and by $\widetilde{\mathcal{N}%
\text{ }}$ the C*-algebra of observables localized in $[0,\infty \lbrack .$
Let $J_{0}$ be the CPT-symmetry of our model. Then we easily have by strong
additivity \cite{Wie2} 
\begin{equation}
J_{0}\widetilde{\mathcal{N}\text{ }}J_{0}\vee \widetilde{\mathcal{N}\text{ }}%
=\mathcal{A}\text{.}
\end{equation}
Let us pass to the weak closure of these algebras w.r.t. the thermal state,
denoted by $\mathcal{M}$ and $\mathcal{N}$ . Under these circumstances $U$
is the modular group associated with $(\mathcal{M},\Omega
_{+}).\,\,\,\,\,\,\,\,\,$By the very definitions \cite{Wie1} we have $(%
\mathcal{N}\subset \mathcal{M},\Omega _{+})$ is a -hsm inclusion. Now $J_{0}%
\widetilde{\mathcal{N}\text{ }}J_{0}\subset \mathcal{N}^{\prime }\cap 
\mathcal{M}$, and it is easy to see that therefore $(\mathcal{N}\subset 
\mathcal{M},\Omega _{+})$ is a standard -hsm inclusion. But we even have $%
\mathcal{M=N}\vee (\mathcal{N}^{\prime }$ $\cap \mathcal{M)}$ , see \cite
{G/L/W}. Now in general , we have due to Borchers result \cite{Bo1}\cite
{Wie1} 
\begin{equation}
\lbrack J_{0},U(a)]=0,
\end{equation}
which easily implies that 
\begin{equation}
A\mapsto <\Omega _{+},J_{0}A^{\ast }J_{0}\Omega _{+}>,\,\,A\in \mathcal{A}
\end{equation}
is again a thermal state of the chiral theory w.r.t. the lightlike
translations. The absence of phase transitions in chiral theories implies
that we can anti-unitarily implement the CPT-operator. Then we immediately
get with $J_{0}\widetilde{\mathcal{N}}J_{0}=\mathcal{N}^{\prime }\cap 
\mathcal{A}$%
\begin{equation}
J_{0}\mathcal{N}J_{0}=\mathcal{N}^{\prime }\cap \mathcal{M}\text{ .}
\end{equation}
This implies that the relative commutant of the observable algebra to a
half-line is the algebra to the opposite half-line.

In case of a -hsm standard inclusion we can always construct out of these
data a chiral vacuum net \cite{Wie2}\cite{G/L/W}. Therefore, within the
above thermal situation we encounter a ``hidden'' vacuum theory. For this,
let us sketch the general construction in the context of this particular
case. One starts by associating the algebra $\mathcal{M}$ to the upper
half-circle. Then exploiting the fact that the modular groups of $\mathcal{M}
$, $\mathcal{N}$, and $\mathcal{M\cap N}^{\prime }$ generate a positive
energy representation of the M\"{o}bius group, we implement the covariance
of the algebra system by defining the various local algebras by mapping $%
\mathcal{M}$ according to the modular representation of the space-time
symmetry. In this way we generate a M\"{o}bius-covariant net.

Let us now compare both theories from a more conceptual point of view.
First, they have the weak closure $\mathcal{M}$ of the quasilocal algebra $%
\mathcal{A}$ in common. Further, the modular theory $U(a)$ to that algebra
w.r.t. $\Omega _{+}$ acts geometrically in both systems, namely as lightlike
translations in the thermal case, and as half-line dilations in the vacuum
case.

Secondly, they have the observable algebras $\mathcal{N}$ and $\mathcal{N}%
^{\prime }$ $\cap \mathcal{M}$ in common, which in the thermal case are
interpreted as the algebras to the two half-lines beginning at origin,
whereas in the vacuum case they are associated to $[1,\infty \lbrack $, and $%
[0,1]$.

Now we notice that the physical symmetry in the thermal case is given by the
lightlike translations, whereas in the vacuum theory we have a natural
conformal symmetry $SL(2,R)/Z_{2}$ behind, where the lightlike translations
of the former theory are equal to the dilations in the latter.

Let us exploit the common structures in order to compare the localizations
in both theories. For this we denote the local algebras by $\mathcal{A}%
_{th}(a,b)$ if we refer to the thermal theory, $\mathcal{A}_{0}(a,b)$ in the
vacuum case. So we can rephrase the above remark by 
\begin{equation}
\mathcal{A}_{th}(0,\infty )=\mathcal{N}=\mathcal{A}_{0}(1,\infty
),\,\,\,\,\,\,\,\,\,\,\mathcal{A}_{th}(-\infty ,0)=\mathcal{N}^{\prime }\cap 
\mathcal{M=A}_{0}(0,1)
\end{equation}
and $U(2\pi t)$ $=\Delta _{\mathcal{M}}^{it}$ where we set the inverse
temperature to be $2\pi .$ (The necessary modifications for the general case
is left to the interested reader.). Then we have 
\begin{eqnarray}
\mathcal{A}_{th}(a,\infty ) &=&Ad\,\,U(a/2\pi )(\mathcal{A}_{th}(0,\infty
))=Ad\,\,U(a/2\pi )(\mathcal{N}) \\
&=&\mathcal{A}_{0}(e^{2\pi a},\infty )  \nonumber
\end{eqnarray}
and similarly 
\begin{equation}
\mathcal{A}_{th}(-\infty ,a)=\mathcal{A}_{0}(0,e^{2\pi a}).
\end{equation}
Now, by the very definition we have 
\begin{equation}
\mathcal{A}_{0}(e^{2\pi a},e^{2\pi b})=\mathcal{A}_{0}(0,e^{2\pi b})\cap 
\mathcal{A}_{0}(e^{2\pi a},\infty )
\end{equation}
and similarly 
\begin{equation}
\mathcal{A}_{th}(a,b)=\mathcal{A}_{th}(-\infty ,b)\cap \mathcal{A}%
_{th}(a,\infty )
\end{equation}
holds due to the strong additivity of the thermal net. Therefore we conclude 
\begin{eqnarray}
\mathcal{A}_{th}(a,b) &=&\mathcal{A}_{th}(-\infty ,b)\cap \mathcal{A}%
_{th}(a,\infty ) \\
&=&\mathcal{A}_{0}(0,e^{2\pi b})\cap \mathcal{A}_{0}(e^{2\pi a},\infty ) 
\nonumber \\
&=&\mathcal{A}_{0}(e^{2\pi a},e^{2\pi b}).  \nonumber
\end{eqnarray}
But now we know, by the very definition, that the conformal group $%
Sl(2R)/Z_{2}$ acts geometrically on the vacuum theory. And due to the above
relation we immediately get the action of that group on the thermal theory.
Therefore, restricting on vacuum localization intervals in the half-line $%
[0,\infty \lbrack $ and to group actions mapping this interval into another
one contained in the half-line, we can transfer this action onto a
geometrical action in the thermal theory. Moreover we see, that there is a
natural space-time structure also on the shadow world i.e. on the thermal
commutant to the quasilocal algebra on which this hidden symmetry naturally
acts. In conclusion, we have encountered a rich hidden symmetry lying behind
the tip of an iceberg, of which the tip was first seen by Borchers and
Yngvason.

In fact using our previous result on the modular origin of the
diffeomorphism for the special case of the Weyl-algebra \cite{S-W1}, it
seems plausible that there is even a much larger infinite dimensional
symmetry of modular origin hidden in this thermal situation.

Before we make some qualitative remarks about hidden SL(2,\textbf{R})
symmetries in the higher dimensional case, let us separately look at d=1+1
massive theories. It is clear that in this case we should use the two
modular inclusions which are obtained by sliding the (right hand) wedge
inside itself ($\mathcal{M}\rightarrow \mathcal{M}_{a_{\pm }})$ by applying
an upper/lower lightlike translation $a_{\pm }$ and forming the relative
commutant (the size of $a_{\pm }$ is irrelevant) 
\begin{equation}
\mathcal{M}(I_{\pm })\equiv \mathcal{M}_{a_{\pm }}^{\prime }\cap \mathcal{M}
\end{equation}
where the notation indicates that the localization of $\mathcal{M}(I_{\pm })$
is thought of as the piece of the upper/lower light ray between the origin
and the endpoint of the $a_{\pm }$ lightlike translation. By viewing this
relative commutant as a lightlike limiting case of a spacetlike shift of $W$
into itself (and using Haag duality), on obtains the interval $I_{\pm }$ as
a limit of a double cone. The net obtained by applying the modular
transformation $\Delta _{M}^{it}$ to the $\mathcal{M}(I_{\pm })$ via its $ad$
action is a chiral net with total algebra 
\begin{equation}
\widetilde{\mathcal{M}}_{\pm }\equiv \bigcup_{t}\Delta _{W}^{it}\left( 
\mathcal{M}(I_{\pm })\right) \subset \mathcal{M}\text{,}
\end{equation}
where we used the hsm standard inclusion $(\mathcal{M}(I_{\pm })\subset $ $%
\widetilde{\mathcal{M}}_{\pm },\Omega )$ for the construction, see the
Appendix below. Since this net is chiral, it cannot create the full space
from the vacuum. Rather the cyclically generated space is a genuine subspace
with projector $P_{\pm }.$ Since the modular group of ($\widetilde{\mathcal{M%
}}_{\pm },\Omega )$ is obtained from restricting $\Delta _{\mathcal{M}}^{it}$
to $P_{\pm }\mathcal{H}$, the projection is associated with a conditional
expectation of the algebra 
\begin{equation}
E_{\pm }(\mathcal{M})=P_{\pm }\mathcal{M}P_{\pm }=\widetilde{\mathcal{M}}%
_{\pm }
\end{equation}
Although the two-dimensional conformal theory lives in the tensor-product
space of the two chiral theories on the upper/lower light ray, the two
chiral components constructed in the present way do not commute with each
other unless one also performs the massless limit m$\rightarrow 0.$ Indeed
since the upper/lower light ray have a relative $\pm $time-like separation,
one does not expect Huygens principle to become effective outside of $m=0.$
In fact a simple straightforward calculation on algebras generated by
massive currents or energy-momentum tensor reveals that the relative
commutator is proportional to a power of the mass but independent of the
space-time coordinates. The mechanism can be illustrated in the simplest way
by looking at the relative commutator of the 1-2 components of the Fermion
field on the upper/lower light ray 
\begin{eqnarray}
\left\{ \psi (x),\bar{\psi}(y)\right\}  &=&\left( i\gamma _{\mu }\partial
^{\mu }-m\right) i\Delta (x-y) \\
i.e.\,\,\left\{ \psi _{1}(u)\psi _{2}^{\ast }(v)\right\}  &\simeq &\pm m 
\nonumber
\end{eqnarray}
where the last relation holds in the limit x,y$\rightarrow $upper/lower
light ray.$\,$So it appears that one has to distinguish between the
tensorproduct algebra $\mathcal{M}(I_{+})\otimes \mathcal{M}(I_{-})$
associated with the full d=1+1 conformal theory (the zero mass limit) and
the combined $\mathcal{M}(I_{+})\vee \mathcal{M}(I_{-})$ subalgebra of $B(%
\mathcal{H}).$ In fact, since one expects the vacuum to be cyclic in $%
\mathcal{H}$ with respect to $\mathcal{M}(I_{+})\vee \mathcal{M}(I_{-}),$
from the shared modular group structure with $\mathcal{M}$ we will take $%
\mathcal{M}=\mathcal{M}(I_{+})\vee \mathcal{M}(I_{-}).\,\,\,$for granted So
modulo a fine point concerning the difference to a tensor product, the two
conformal symmetries $SL(2,\mathbf{R})/Z_{2}$ of the upper/lower light ray
reductions are in fact inherited as hidden symmetries by the massive theory.
The light ray reduction of the massive theory, although being a chiral
conformal theory in its own right, can be transmuted into the original
massive theory by adjoining the opposite light cone translation $U_{-}(a)$
i.e. by extending $M(I_{+})$ to $alg\left\{ M(I_{+}),U_{-}(a)\right\} $ \cite
{S-W2}$.$ One would expect that the relation between the thermal version and
the groundstate representation of the massive theory goes through the light
ray reduction. In addition to the proper conformal transformation which are
already hidden in the massive theory there should be a ``thermal hiding''
for all transformations except the translations.

\section{Qualitative Remarks about Higher Dimensions}

In a previous paper we have shown how for $d\geq 1+2$ one may use the
situation of modular intersections to come to hidden symmetries \cite{S-W1}.
In higher dimensions one expects that in general a conversion of thermal
nets into spatially extended ground state nets, if possible at all, will
involve $SL(2,R)/Z_{2}$ symmetries of modular origin with positive energy
translation operators i.e.symmetries which (since a massive theory does not
possess such pointwise acting symmetries) have hidden pieces already in
their ground state realization.

For this purpose we introduce the notion of a ``conformal core'' with the
``blow-up'' property. Let us start with a d=1+1 theory defined in terms of
algebraic net data i.e. a coherent map of space-time regions into von
Neumann algebras. Such theories have two light-like translations $U_{\pm
}(a) $ with positive generators. We use the right upper light ray
translation $U_{+}$ in order to construct a modular inclusion by sliding the
standard wedge $W$ at the origin into itself $W\rightarrow W_{+}\subset W$
using $U_{+}(a=1)$ \cite{S-W2}$.$ Consider the modular inclusion defined in
terms of the relative commutant of $\mathcal{M}_{+}\equiv U(1)\mathcal{A}%
(W)U^{*}(1)$ in $\mathcal{M}=\mathcal{A}(W)$ 
\begin{equation}
(\mathcal{M}_{+}^{\prime }\cap \mathcal{M}\subset \mathcal{M},\Omega 
\mathcal{)}
\end{equation}
This inclusion leads to a full-fledged chiral conformal theory on the line
(the light ray reduction) which generates from the vacuum its own Hilbert
space which is a genuine subspace of the original space 
\begin{eqnarray}
\mathcal{A}_{\pm }^{>} &\equiv &\bigcup_{t}\Delta _{W}^{it}\left( \mathcal{M}%
_{\pm }^{\prime }\cap \mathcal{M}\right) \subset \mathcal{M}\equiv \mathcal{A%
}(W) \\
\overline{\mathcal{A}_{\pm }^{>}\Omega } &=&H_{\pm }\subset H=\overline{%
\mathcal{M}\Omega }  \nonumber \\
\mathcal{A}_{\pm } &=&\mathcal{A}_{\pm }^{>}\vee J\mathcal{A}_{\pm }^{>}J 
\nonumber
\end{eqnarray}
Note that the full conformal invariance is only realized on the reduced
space $H_{\pm }$

Since the modular theories of the combined upper/lower light ray reductions $%
\mathcal{A}_{+}^{>}\vee \mathcal{A}_{-}^{>}$ defines a 2-dim. net which
lives on a bigger Hilbert space than the reduced one and since this combined
algebra has the same modular theory and the same light-like translations as $%
\mathcal{M}$ there exists according to Takesaki a conditional expectation
from $\mathcal{M}$ to $\mathcal{A}_{+}^{>}\vee \mathcal{A}_{-}^{>}$ and the
two algebras are either identical (iff the cyclically generated Hilbert
spaces are the same) or the $\mathcal{A}_{+}^{>}\vee \mathcal{A}_{-}^{>}$
-algebra is obtain from $\mathcal{M}$ as a fixed point algebra under the
action of an internal symmetry. A more efficient way to reconstruct the
original massive net from its light ray reduction is the following
``blow-up'' of a light ray reduction, say $\mathcal{A}_{+},$ by use of the
opposite positive ($a>0$) light-like translation 
\begin{equation}
\mathcal{B}_{+}\equiv \bigvee_{a>0}alg\left\{ A_{+},U_{-}(a)\,\right\}
\end{equation}

Intuitively one expects that $\mathcal{M}$, $\mathcal{A}_{+}^{>}\vee 
\mathcal{A}_{-}^{>}$ and $\mathcal{B}_{+},\mathcal{B}_{-}$ are all
identical, but without further detailed investigations, which go beyond the
identity of their modular structure, we cannot decide whether there are
differences due to different internal symmetries.

This light ray reduction supplemented by the blow-up idea can be generalized
to higher dimensions, where however it becomes more tricky. Let us explain
the situation for d=2+1, using the notation of \cite{S-W1}:

$U_{12,13}$ for the Galilean ``translations'' inside the Lorentz group,
which in d=2+1 exhausts the isotropy (``little'') group to $l_{1}.$

$U_{l_{1}}$ for the lightlike translations along $l_{1},$ $\mathcal{A}%
([l_{1},l_{2}])$ for the wedge algebra to $W[l_{1},l_{2}]$

With this notational matter out of our way, we present now two interesting
proposals for the definition of candidates for a ``chiral conformal core''
associated to a lightlike direction. The first one leads to a construction
which by fiat is independent of the choice of the wedge and only depends on
the light ray $l_{1}$. Start from

\begin{equation}
Ad\,\,\,U_{l_{1}}(1)(\mathcal{A}([l_{1},l_{2}])\subset \mathcal{A}%
([l_{1},l_{2}]).
\end{equation}

This defines a conformal theory, see the appendix. The resulting
translations, namely $U_{1},$ commutes with the Galilean ``translations''.
We define:

\begin{equation}
\cap _{\lambda \in R}Ad\,\,U_{l_{1}l_{2},l_{1}l_{3}}(\lambda
)(Ad\,\,\,U_{l_{1}}(1)(\mathcal{A}([l_{1},l_{2}]))\subset \mathcal{A}%
([l_{1},l_{2}]).  \label{alg}
\end{equation}

This gives a modular standard inclusion in a canonical way, see appendix.
This proposal has the advantage to be covariant under the action of the
isotropy:$\,\,\,\,\,\,\,\,$%
\begin{eqnarray}
&&U_{l_{1}l_{2},l_{1}l_{3}}(\mu )\cap _{\lambda \in
R}Ad\,\,U_{l_{1}l_{2},l_{1}l_{3}}(\lambda )(Ad\,\,\,U_{l_{1}}(1)(\mathcal{A}%
([l_{1},l_{2}])) \\
&=&Ad\,\,\,U_{l_{1}}(1)(\cap _{\lambda \in
R}Ad\,\,U_{l_{1}l_{2},l_{1}l_{3}}(\lambda )U_{l_{1}l_{2},l_{1}l_{3}}(\mu )((%
\mathcal{A}([l_{1},l_{2}])))  \nonumber \\
&\subset &U_{l_{1}l_{2},l_{1}l_{3}}(\mu )(\mathcal{A}([l_{1},l_{2}])). 
\nonumber
\end{eqnarray}
where the Galilean translation $U_{l_{1}l_{2},l_{1}l_{3}}(\mu )$ simply
turns $W[l_{1},l_{2}]$ into a $W[l_{1},l_{2}^{\prime }].$

The drawback of this construction could be that it is empty, as we will
argue in the sequel.

An equivalent construction based on the same intuition starts with the
modular inclusion

\begin{equation}
\mathcal{A}([l_{1},l_{2}])\cap \mathcal{A}([l_{1},l_{3}]))\subset \mathcal{A}%
([l_{1},l_{2}])
\end{equation}

and suggests to represent the algebra (\ref{alg}) as a limiting intersection
algebra: 
\begin{equation}
\lim_{\lambda \rightarrow \infty }Ad\,\,U_{l_{1}l_{2},l_{1}l_{3}}(\lambda )(%
\mathcal{A}([l_{1},l_{2}])\cap \mathcal{A}([l_{1},l_{3}]))
\end{equation}
and this seems to be $\mathbf{C}\cdot 1,$ a multiple of the identity.

Let us therefore mention another possible construction which starts with the
modular inclusion:

\begin{equation}
Ad\,\,\,U_{l_{1}}(1)(\mathcal{A}([l_{1},l_{2}])\cap \mathcal{A}%
([l_{1},l_{3}]))\subset \mathcal{A}([l_{1},l_{2}])
\end{equation}

Again intuitively this might be interpreted as a conformal core to the light
ray $l_{1}.$ This definition depends on the lightlike directions, but it
does so in a covariant way since $\,U_{l_{1}l_{2},l_{1}l_{3}}(\lambda )$
commutes with $U_{l_{1}}$ and therefore the action of the isotropy group is
computable. Moreover this modular inclusion is only invariant under the
lighlike translations $U_{l_{1}}$and not under\thinspace the transversal
translations.

In a generic curved space-time situation with a bifurcated Killing horizon
and in the absence of additional symmetries, one still can apply the methods
of LQP \cite{S/S-V} and associate a chiral conformal theory as was shown
recently in a remarkable paper by Guido, Longo, Roberts and Verch \cite
{G/L/R/V}. In that case the conformal theory is not associated to a
particular light ray, but rather is induced via a restriction of the theory
onto the horizon.

Before we apply the achieved results to our main theme, namely how to
convert heat bath temperature into Unruh temperature (or how to pass from
the heat bath ``shadow world'' of the thermal commutant into new space-time
behind a horizon), we cannot resist to make some comments about a
fascinating but speculative connection of the blow-up picture of the chiral
core with another speculative problem which presently is attracting the
attention of many theoreticians: the problem of quantum entropy of algebras
with a horizon or (in the context of LQP) a ``quantum localization
entropy''. The blow-up picture tell us that in the case of the wedge
geometry we can represent the original wedge algebra $\mathcal{M}$ by the
algebra generated by the chiral conformal algebra augmented with certain
symmetry generators which do the blowing up (principally longitudinal and
transverse translations). One would expect that for the counting of degrees
of freedom the Poincar\'{e} generators do not contribute. If we now use our
physical imagination for an attempt to understand this picture beyond the
wedge also for the rotationally symmetric double cone (in Minkowski space),
and take notice that if the higher dimensional theory would be conformally
invariant (zero mass), then there is a well-known transformation which
carries the previous wedge situation into the double cone, then we obtain
indeed a very attractive picture. The more convincing part is the blow-up
representation of the higher dimensional conformal algebra in terms of a
chiral core algebra on the surface (plus symmetry generators of the higher
dimensional theory which do not participate in the degree of freedom
counting). The second, less rigorous step, is the idea that the unknown
``fuzzy'' modular theory of the massive double cone algebra is
asymptotically (near the horizon) equal to the geometric modular situation
of the conformal double cone theory. However to convert this ``holographic
degree of freedom picture'' into a Bekenstein-Hawking-like formula for
entropy expressed in terms of data of the chiral conformal core theory (e.g.
the structure constants of W-algebras)  more needs to be done. There is
finally the problem of defining what we mean by localization entropy in the
chiral theory. The local algebras, whether chiral or higher dimensional, are
known to be of von Neumann hyperfinite type III$_{1}$ which would lead to
diverging (undefined) von Neumann entropies. As physicists we would try to
regularize the situation. Indeed there is a natural way to do this, which is
related to the phase space behavior of QFT. The latter is different from the
behavior obtained by the box quantization at fixed time in the sense that
the density based on the correct localization concept is bigger
(``nuclear'') than for the box case (finite) \cite{Ha}, a fact overlooked in
most relativistic entropy discussions. Closely related to the nuclearity
property is the so-called split property which suggests to define a kind of
fuzzy localized type I algebra. This algebra has its support in the double
cone plus a ``collar'' around it, so that there is a bigger double cone
containing both which the given double cone and the collar \cite{Ha}. A type
I algebra, unlike hyperfinite type III$_{1},$ has no principle obstruction
against the existence of a von Neumann entropy. Such an entropy takes into
account the quantum ``entaglement'' between the inside of the smaller and
the ouside of the bigger double cone across the collar. With vanishing size
of the collar the entropy diverges which corresponds to the infinite (but
still nuclear) correct degree of freedom counting. Filling the double cone
with different kind of matter, the strength of divergence is expected to be
the same but the leading coefficients vary. The missing and difficult part
of the discussion is the use of the vacuum state, because the modular
simplicity of the doubly localized situation with the collar has a
geometrically simple modular theory only with respect to another state \cite
{S-W1} whose Connes cocycle relation to the vacuum has nit been studied.  An
educated guess suggests that the divergence is logarithmic; also the claims
which appeared in the literature that for minimal models the coefficient is
proportional to the to the central charge constant $c,$ are certainly not in
contradiction with our blow-up picture of chiral cores and very suggestive
indeed if one looks at the aforementioned relation of the two states and the
role of the central term of the diffeomorphism group in the construction of
the geometric modular state. Note that these arguments, if they withstand
furthergoing detailed analysis, would interpret the holographic behavior as
a \textit{generic property of nonperturbative local quantum physics} and not
as a particular behavior of special global properties\footnote{%
The present authors are aware of surprizing recent observations on entropy
within the setting of string theories. Either string theories have locality
properties (not known), in which case they are special LQP's; or they are
something nonlocal in which case, as a result of lack of physical
interpretability which for all important properties (scattering theory,
superselected charges, statistics...) totally hinges on locality, they are
remarkable mathematical (but physically nonunderstood) observations. The
fruitful conceptual curiosity from the first half of this century which was
crucial to resolve such such interesting situatios unfortunately seems to
have been lost with the invention of the very string theory which led to
those observations.} (topological field theories, gauge theories etc.). This
yet speculative scenario build on modular theory has a some resemblance with
Wald's recent more conservative ideas on entropy \cite{Wa} and less so with
string theory; although the results (but certainly not the physical
concepts) may be similar.

The reader may find an earlier account of this picture on holographic
reduction of degrees of freedom from the viewpoint of LQP in section of a
book manuscript draft by one of the authors \cite{Schr1}; the present
blow-up mechanism of chiral light ray reduction lends considerably more
credibility to those earlier remarks. There is a history and a long list of
references on light cone holography in black hole physics by a variety of
methods which seems to be based on quite different ideas than those of this
work \cite{Ca}. We hope to return with more results to this interesting and
conceptually important entropy problem.

Returning now to the thermal theme of this paper, the qualitative idea of
generalizing our discussion of the Borchers-Yngvason model to higher
dimensions is quite simple: do the construction of the previous section on a
conformal core and then use the blow-up construction using the longitudinal
and transversal translations once in the ground state and once in the
thermal representation. Whether this really works in higher dimensional
model cases, still remains to be seen.

Finally we comment on the problem of spontaneously broken symmetries. It is
well-known that the change of temperature is often accompanied by a
transition of phase related to a change in global symmetry. The possibility
of a modular transmutation of a thermal into a ground state theory (with
spatial extension behind a horizon) creates all kinds of curious consistency
problems.

Unfortunately the better understood chiral conformal theories do not allow
for spontaneous symmetry breaking. The only exceptional case is the thermal
collapse of supersymmetry \cite{B-O}. Since the ``collapse mechanism''
originates from the impossibility of annihilating a faithful state (any
thermal state is faithful) on a C$^{*}$-algebra by an antiderivation, it is
independent of space-time dimensions and holds in particular in chiral
theories. Although it does not serve as a an illustration of spontaneous
symmetry breaking, its thermal aspects are interesting in their own right.
Assuming that our transmutation mechanism also works for general chiral
models including supersymmetric ones, and confronting it with the collapse
mechanism one finds a somewhat curious situation whose only resolution seems
to be that the (Unruh) thermal restriction wrecks the action of the
antiderivative on only one side of the horizon and in this way ``explains''
the violent collapse. Whether this results allow for a deeper understanding
of supersymmetry or only adds to the growing suspicion that this symmetry is
of an accidental nature \cite{Schr2}, remains to be seen.

\section{Concluding Remarks}

Although the fully pointwise geometric symmetries which are well-known from
the quantization approach to relativistic QFT are, with the exception of
chiral conformal QFT, restricted to finite dimensional automorphisms as
Poincar\'{e} symmetry and (only for zero mass) conformal symmetry, the
modular structure of LQP opens the gates for a vast variety of infinite
dimensional groups. One may either vary the states and consider the modular
groups generated by wedge- and double cone-algebras or one may investigate
the modular inclusions/intersections generated from wedges with respect to
natural reference states od curved sace-time or the vacuum state in case of
Minkowski space-time. As we argued in \cite{S-W1}, we expect in the first
case to obtain an (infinite dimensional) ``hidden'' analogue of the chiral
diffeomorphism group. This is based on the fact that the chiral
diffeomorphism group can be build up from infinitely many ``M\"{o}bius
layers'' by lifting the vacuum M\"{o}bius group with a covering
transformation and by realizing that each such n$^{th}$ layer lifted M\"{o}%
bius group\footnote{%
These lifted Moebius transformations $Moeb_{n}$ are called
``quasisymmetric'' of order $n$ in the mathematical literature.} is the
modular group of a pair ($\mathcal{A}(I_{1})\vee ...\vee \mathcal{A}%
(I_{n}),\Omega _{n}).$ The von Neumann algebra n this case is n-fold
localized and the ``lifted vacua'' $\Omega _{n}$ are suitably chosen states
such that the endpoints of the $I_{k},$ $k=1...n$ are fixed points of the n$%
^{th}$ layer M\"{o}bius transformation and $\Omega _{n}$ is its unique
invariant state \cite{S-W1}. The higher dimensional analogues of this
construction beyond the Poincar\'{e} resp. conformal group are expected to
be fuzzy (nonlocal) independent of the chosen state; in fact according to a
conjecture of Fredenhagen their infinitesimal action on test functions
should be described by pseudo-differential operators which are at most
asymptotically geometric near the horizon i.e. the space-time border of the
localization region). In the present paper we have studied the partially
hidden symmetry groups associated with thermal modular inclusions in chiral
models. Even though in this case we do find geometrical aspects, these
symmetries are not implemented by pointlike transformations in the
underlying space-time and hence are hidden in the quantization approach. In
addition the concepts of ``chiral core'' and its ``blow-up property'' which
led us to a holographic degrees of freedom picture and gave some nice ideas
about quantum entropy is obtained by the extremely noncommutative modular
theory. It is a generic property of local quantum theory and cannot be seen
by quantization procedures (and therefore is e.g. not related to
differential geometric properties of Chern-Simons actions nor to gauge
theory). The modular method based on real time noncommutative LQP seems to
develop into a viable alternative to the euclidean functional integral
approach to QFT \cite{S-W2}.

Acknowledgments: One of the authors (B.S.) thanks J. Yngvason, D. Buchholz
and R. Verch for a helpful correspondence.

\textbf{Appendix}

\textbf{Canonical construction which underlies the calculation of the chiral
core}

Let $(\mathcal{N\subset M},\Omega )$ hsm with non trivial relative
commutant. Then look at the subspace $\overline{(\mathcal{N}^{\prime }\cap 
\mathcal{M)}\Omega }\subset H.$ The modular groups to $\mathcal{N}$ and $%
\mathcal{M}$ leave invariant this subspace: $\Delta _{\mathcal{M}}^{it}$
maps $\mathcal{N}^{\prime }\cap \mathcal{M}$ into itself by hsm for say
positive $t$. Look at the orthogonal complement of $\overline{(\mathcal{N}%
^{\prime }\cap \mathcal{M)}\Omega }\subset H.$ This orthogonal complement is
mapped into itself by $\Delta _{\mathcal{M}}^{it}$ for positive $t.$ Let $%
\psi $ be in that subspace, then 
\begin{equation}
\left\langle \psi ,\Delta _{\mathcal{M}}^{it}(\mathcal{N}^{\prime }\cap 
\mathcal{M})\Omega \right\rangle =0\,\,for\,\,t>0.
\end{equation}

$\,$Analyticity in $t$ then gives the vanishing for all $t.$

Due to Takesaki theorem we can restrict $\mathcal{M}$ to $\overline{(%
\mathcal{N}^{\prime }\cap \mathcal{M)}\Omega }$ using a conditional
expectation to this subspace. Then 
\[
(\mathcal{N}^{\prime }\cap \mathcal{M)\subset M}|_{\overline{(\mathcal{N}%
^{\prime }\cap \mathcal{M)}\Omega }} 
\]
is a hsm on the subspace defined above. $\mathcal{N}$ also restricts to that
subspace and this restriction is obviously in the relative commutant of $(%
\mathcal{N}^{\prime }\cap \mathcal{M)\subset M}|_{\overline{(\mathcal{N}%
^{\prime }\cap \mathcal{M)}\Omega }}$ . Moreover using arguments as above it
is easy to see that the restriction is cyclic w.r.t.$\Omega $ on this
subspace. Therefore we arrive at a hsm standard inclusion 
\begin{equation}
(\mathcal{N}|_{\overline{(\mathcal{N}^{\prime }\cap \mathcal{M)}\Omega }%
}\subset \mathcal{M}|_{\overline{(\mathcal{N}^{\prime }\cap \mathcal{M)}%
\Omega }},\Omega )
\end{equation}
$.$

\end{document}